\documentclass[journal]{IEEEtran}
\usepackage{amssymb}

\usepackage{subfigure}
\usepackage{graphicx}
\usepackage{enumerate}
\usepackage{amsmath}
\usepackage{amsfonts}
\usepackage{algorithm}
\usepackage{algpseudocode}
\usepackage{url}
\usepackage{epstopdf}
\usepackage{multirow}
\usepackage{authblk}
\usepackage{subfigure}
\usepackage{xcolor}
\usepackage{mathrsfs}

\usepackage{verbatim}

\newtheorem{theorem}{Theorem}[section]
\newtheorem{example}[theorem]{Example}
\newtheorem{corollary}[theorem]{Corollary}
\newtheorem{proposition}[theorem]{Proposition}
\newtheorem{definition}[theorem]{Definition}
\newtheorem{lemma}[theorem]{Lemma}
\newcommand{\prof}{\begin{IEEEproof}}
\newcommand{\eprof}{\end{IEEEproof}}

\newcommand{\prop}{\begin{proposition}}
\newcommand{\eprop}{\end{proposition}}
\newcommand{\them}{\begin{theorem}}
\newcommand{\ethem}{\end{theorem}}
\newcommand{\dfn}{\begin{definition}}
\newcommand{\edfn}{\end{definition}}
\newcommand{\exm}{\begin{example}}
\newcommand{\eexm}{\end{example}}
\newcommand{\coro}{\begin{corollary}}
\newcommand{\ecoro}{\end{corollary}}
\newcommand{\lem}{\begin{lemma}}
\newcommand{\elem}{\end{lemma}}

\newcommand{\eps}{\varepsilon}

\ifCLASSINFOpdf
\else
\fi
\begin{document}
%
\title{Verification of Detectability in Petri Nets Using Verifier Nets}
%
%
%

\author{Hao~Lan, 
        Yin~Tong, ~\IEEEmembership{Member,~IEEE}
        Carla~Seatzu, \IEEEmembership{Senior Member,~IEEE}
        and~Jin~Guo
\thanks{H. Lan, Y. Tong (Corresponding Author), and Jin Guo are with the School of Information Science and Technology, Southwest Jiaotong University, Chengdu 611756, China
        {\tt\small haolan@my.swjtu.edu.cn; yintong@swjtu.edu.cn}}
\thanks{C. Seaztu is with the Department of Electrical and Electronic Engineering,
University of Cagliari, 09123 Cagliari, Italy {\tt\small seatzu@diee.unica.it}}}%

%
%

\markboth{Journal of \LaTeX\ Class Files,~Vol.~14, No.~8, August~2015}%
{Shell \MakeLowercase{\textit{et al.}}: Bare Demo of IEEEtran.cls for IEEE Journals}
%



\maketitle

\begin{abstract}
Detectability describes the property of a system whose current and the subsequent states can be uniquely determined after a finite number of observations. In this paper, we developed a novel approach to verifying strong detectability and periodically strong detectability of bounded labeled Petri nets. Our approach is based on the analysis of the basis reachability graph of a special Petri net, called Verifier Net, that is built from the Petri net model of the given system. Without computing the whole reachability space and without enumerating all the markings, the proposed approaches are more efficient.
\end{abstract}


%
\IEEEpeerreviewmaketitle

\section{Introduction}\label{sec:intro}
In recent years detectability has drawn a lot of attention from researchers in the discrete event system (DES) community. This property characterizes the ability of a system to determine the current and the subsequent states of the system after the observation of a finite number of events.

Detectability has been studied earlier in DES under the name of observability \cite{ramadge1986observability,ozveren1990observability,giua2002observability}. The observability of the current state and initial state are discussed in \cite{ramadge1986observability}, and whether the current state can be determined periodically is investigated in \cite{ozveren1990observability}.
The property of detectability in DESs has been studied systematically in the literature \cite{shu2007detectability,shu2011generalized,zhang2017problem,masopust2018complexity, keroglou2015detectability}.
The notion of detectability was first proposed and studied in \cite{shu2007detectability} in the deterministic finite automaton framework based on the assumption that the states and the events are partially observable. Shu et al. \cite{shu2007detectability} defined four types of detectability: strong detectability, weak detectability, strong periodic detectability, and weak periodic detectability. And the four types of detectability are verified by an approach whose complexity is exponential with respect to the number of states of the system. Polynomial algorithms to check strong detectability and strong periodic detectability of an automaton (called detector) have been proposed in \cite{shu2011generalized}. While checking weak detectability and weak periodic detectability is proved to be PSPACE-complete and that PSPACE-hardness \cite{zhang2017problem}, even for a very restricted type of automata \cite{masopust2018complexity}. The notation of detectability is also extended to delayed DESs \cite{shu2013delayed}, modular DESs \cite{yin2017verification} and stochastic DESs \cite{keroglou2015detectability,yin2017initial}, and the enforcement of the detectability is proposed in \cite{shu2013online,yin2016uniform}.

Petri nets are widely used to model many classes of concurrent systems, some problems such as supervisory control \cite{ma2015design}, fault diagnosis \cite{cabasino2011discrete}, opacity \cite{tong2017verification}, etc. can be solved more efficiently in Petri nets.
The detectability of unlabeled Petri nets was proposed by Giua and Seatzu \cite{giua2002observability}, including marking observability and strong marking observability.
In \cite{masopust2018deciding}, the authors extend strong detectability and weak detectability in DESs to labeled Petri nets. Strong detectability is proved to be decidable and checking the property is EXPSPACE-hard, while weak detectability is proved to be undecidable.
In our previous work \cite{tong2019verification}, we first extend the four detectability to labeled Petri nets and then based on the notion of \emph{basis markings} efficient approaches to verifying the four detectability are proposed. However, the method in \cite{tong2019verification} requires the construction of an observer of the \emph{basis reachability graph} (BRG) of the LPN system. Since in the worst case, the complexity of constructing the observer is exponential to the number of states of the BRG. Thus, it is important to search for more efficient algorithms for checking detectability in labeled Petri nets.

In this paper, we develop a method to check strong detectability and periodically strong detectability with lower complexity. The method is based on the construction of a new tool, called "verifier net", which was first proposed in \cite{cabasino2012new} for verification of diagnosability. The verifier net is a special labeled Petri net, that is built from the original LPN system. Then, we present necessary and sufficient conditions for the strong detectability and periodically strong detectability, by analyzing the BRG of the verifier net. Therefore, the construction of the observer is avoided. The efficiency and effectiveness of the proposed approach is shown by comparing it with our previous method \cite{tong2019verification} in Section~\ref{subsec:complexity}.

We assume that both the structure and the initial marking of the Petri net are known, and the system¡¯s evolution is only partially observed. Our work is related to several works on state estimation of Petri nets \cite{masopust2018deciding,cabasino2012new}. In particular, it is closely related to the work of Masopust and Yin \cite{masopust2018deciding} who proposed a \emph{twin-plant} construction algorithm to verify the strong detectability. However, in this paper, the difference here is that 1) our verifier net is a special labeled Petri net, and we modified the labeling function of the Verifier net for detectability in the construction algorithm, which is different from \cite{cabasino2012new}; 2) we construct the BRG of the verifier net which is no need to enumerate all the markings; 3) we use this approach not only to check the strong detectability but also the periodically strong detectability, and we also explained why this approach for weak detectability and periodically weak detectability is more complex.


The rest of the paper is organized as follows. In Section~\ref{sec:pre}, backgrounds on finite automata, labeled Petri nets, basis markings and the definition of four detectabilities are recalled. The property of the verifier net and its BRG is proposed in Section~\ref{sec:ver}. In Section~\ref{sec:verfication}, the efficient approaches to verifying the strong detectability, periodically strong detectability are presented. Finally, the paper is concluded and the future work is summarized. 
\section{Preliminaries and Background}\label{sec:pre}
In this section we recall the formalism used in the paper and some results on state estimation in Petri nets. For more details, we refer to \cite{cabasino2011discrete,murata1989petri,cassandras2009introduction}.

\subsection{Automata}\label{subsec:auto}
A \emph{nondeterministic finite automaton} (NFA) is a 4-tuple $A=(X, E, f, x_0)$, where $X$ is the finite \emph{set of states}, $E$ is the finite \emph{set of events}, $f: X\times E\rightarrow 2^X$ is the (partial) \emph{transition relation}, and $x_0\in X$ is the \emph{initial state}. The transition relation $f$ can be extended to $f:X\times E^*\rightarrow 2^X$ in a standard manner. Given an event sequence $w\in E^*$, if $f(x_0,w)$ is defined in $A$, $f(x_0,w)$ is the set of states reached in $A$ from $x_0$ with $w$ occurring.

Given an NFA, its equivalent DFA, called \emph{observer}, can be constructed following the procedure in Section~2.3.4 of \cite{cassandras2009introduction}. Each state of the observer is a set of states from $X$ that the NFA may be in after an event sequence occurring. Thus, the complexity, in the worst case, of computing the observer is ${\cal O}(2^n)$, where $n$ is the number of states of $A$.

\subsection{Petri Nets}\label{subsec:pn}
A \emph{Petri net} is a structure $N=(P,T,Pre,Post)$, where $P$ is a set of $m$ \emph{places}, graphically represented by circles; $T$ is a set of $n$ \emph{transitions}, graphically represented by bars; $Pre:P\times T\rightarrow\mathbb{N}$ and $Post:P\times T\rightarrow\mathbb{N}$ are the \emph{pre-} and \emph{post-incidence functions} that specify the arcs directed from places to transitions, and vice versa. The incidence matrix of a net is denoted by $C=Post-Pre$. A Petri net is said to be \emph{acyclic} if there are no oriented cycles.

A \emph{marking} is a vector $M:P\rightarrow \mathbb{N}$ that assigns to each place a non-negative integer number of tokens, graphically represented by black dots. The marking of place $p$ is denoted by $M(p)$. A marking is also denoted as $M=\sum_{p\in P}M(p)\cdot p$. A \emph{Petri net system} $\langle N,M_0\rangle$ is a net $N$ with \emph{initial marking} $M_0$.

A transition $t$ is \emph{enabled} at marking $M$ if $M\geq Pre(\cdot,t)$ and may fire yielding a new marking $M'=M+C(\cdot,t)$. We write $M[\sigma\rangle$ to denote that the sequence of transitions $\sigma=t_{j1}\cdots t_{jk}$ is enabled at $M$, and $M[\sigma\rangle M'$ to denote that the firing of $\sigma$ yields $M'$. The set of all enabled transition sequences in $N$ from marking $M$ is $L(N,M)=\{\sigma\in T^*| M[\sigma\rangle \}$. Given a sequence $\sigma\in T^*$, the function $\pi:T^*\rightarrow \mathbb{N}^n$ associates with $\sigma$ the Parikh vector $y=\pi(\sigma)\in\mathbb{N}^n$, i.e., $y(t)=k$ if transition $t$ appears $k$ times in $\sigma$. Given a sequence of transitions $\sigma\in T^*$, its \emph{prefix}, denoted as $\sigma'\preceq \sigma$, is a string such that $\exists \sigma''\in T^*:\sigma'\sigma''=\sigma$. The \emph{length} of $\sigma$ is denoted by $|\sigma|$.

A marking $M$ is \emph{reachable} in $\langle N,M_0\rangle$ if there exists a sequence $\sigma$ such that $M_0[\sigma\rangle M$. The set of all markings reachable from $M_0$ defines the \emph{reachability set} of $\langle N,M_0\rangle$, denoted by $R(N,M_0)$. A Petri net system is \emph{bounded} if there exists a non-negative integer $k \in \mathbb{N}$ such that for any place $p \in P$ and any reachable marking $M \in R(N,M_0)$, $M(p)\leq k$ holds.

A \emph{labeled Petri net} (LPN) is a 4-tuple $G=(N,M_0,\allowbreak E,\ell)$, where $\langle N,M_0\rangle$ is a Petri net system, $E$ is the \emph{alphabet} (a set of labels) and $\ell:T\rightarrow E\cup\{\eps\}$ is the \emph{labeling function} that assigns to each transition $t\in T$ either a symbol from $E$ or the empty word $\eps$. Therefore, the set of transitions can be partitioned into two disjoint sets $T=T_o\cup T_u$, where $T_o=\{t\in T|\ell(t)\in E\}$ is the set of observable transitions and $T_u=T\setminus T_o=\{t\in T|\ell(t)=\eps\}$ is the set of unobservable transitions.
The labeling function can be extended to sequences $\ell: T^*\rightarrow E^*$ as $\ell(\sigma t)=\ell(\sigma)\ell(t)$ with $\sigma\in T^*$ and $t\in T$.
The set of \emph{language generated by} an LPN $G$ is denoted as ${\cal L}(G)=\{w\in E^*|\exists \sigma\in L(N,M_0):w=\ell(\sigma)\}$.
Let $w\in {\cal L}(G)$ be an observed word. We define ${\cal C}(w)=\{M\in \mathbb{N}^m|\exists \sigma\in L(N,M_0):M_0[\sigma\rangle M, \ell(\sigma)=w\}$ as the set of markings \emph{consistent} with $w$. Markings in ${\cal C}(w)$ are \emph{confusable} with each other as when $w$ is observed it is confused which marking in ${\cal C}(w)$ is the current marking of the system.

Given an LPN $G=(N,M_0,\allowbreak E,\ell)$ and the set of unobservable transitions $T_u$, the \emph{$T_u$-induced subnet} $N'=(P,T',\allowbreak Pre',Post')$ of $N$, is the net resulting by removing all transitions in $T\setminus T_u$ from $N$, where $Pre'$ and $Post'$ are the restriction of $Pre$, $Post$ to $T_u$, respectively. The incidence matrix of the $T_u$-induced subnet is denoted by $C_u=Post'-Pre'$.

\subsection{Basis Markings}\label{sec:basis}
In this subsection we recall some results on state estimation using basis markings proposed in \cite{cabasino2011discrete,ziyue2017basis}.

\dfn\label{def:exp}
Given a marking $M$ and an observable transition $t\in T_o$, we define $$\Sigma(M,t)=\{\sigma\in T^*_u|M[\sigma\rangle M',M'\geq Pre(\cdot,t)\}$$ as the set of \emph{explanations} of $t$ at $M$ and $Y(M,t)=\{y_u\in \mathbb{N}^{n_u}|\exists \sigma\in \Sigma(M,t):y_u=\pi(\sigma)\}$ the set of \emph{$e$-vectors}. \hfill $\diamond$
\edfn

Thus $\Sigma(M,t)$ is the set of unobservable sequences whose firing at $M$ enables $t$. Among all the explanations, to provide a compact representation of the reachability set we are interested in finding the minimal ones, i.e., the ones whose firing vector is minimal.

\dfn\label{def:minexp}
Given a marking $M$ and an observable transition $t\in T_o$, we define
$$\Sigma_{min}(M,t)=\{\sigma\in \Sigma(M,t)|\nexists \sigma'\in \Sigma(M,t):\pi(\sigma')\lneqq\pi(\sigma)\}$$
as the set of \emph{minimal explanations} of $t$ at $M$ and $Y_{min}(M,t)=\{y_u\in \mathbb{N}^{n_u}|\exists \sigma\in \Sigma_{min}(M,t):y_u=\pi(\sigma)\}$ as the corresponding set of \emph{minimal $e$-vectors}. \hfill $\diamond$
\edfn

Many approaches can be applied to computing $Y_{min}(M,t)$. In particular, when the $T_u$-induced subnet is acyclic the approach proposed by Cabasino \emph{et al}. \cite{cabasino2011discrete} only requires algebraic manipulations. Note that since a given place may have two or more unobservable input transitions, i.e., the $T_u$-induced subnet is not backward conflict free, the set of minimal explanations is not necessarily a singleton.

\dfn\label{def:basisM}
Given an LPN system $G=(N,M_0,E,\ell)$ whose $T_u$-induced subnet is acyclic, its \emph{basis marking set} ${\cal M}_b$ is defined as follows:
\begin{itemize}
  \item $M_0\in {\cal M}_b$;
  \item If $M\in {\cal M}_b$, then $\forall t\in T_o, y_u\in Y_{min}(M,t)$,
  $$M'=M+C(\cdot,t)+C_u\cdot y_u \Rightarrow M'\in {\cal M}_b.$$
\end{itemize}
A marking $M_b\in {\cal M}_b$ is called a \emph{basis marking} of $G$.
\edfn

The set of basis markings contains the initial marking and all other markings that are reachable from a basis marking by firing a sequence $\sigma_u t$, where $t\in T_o$ is an observable transition and $\sigma_u$ is a minimal explanation of $t$ at $M$. Clearly, ${\cal M}_b\subseteq R(N,M_0)$, and in practical cases the number of
basis markings is much smaller than the number of reachable markings\cite{cabasino2011discrete,ziyue2017basis}. We denote ${\cal C}_b(w)={\cal M}_b\cap {\cal C}(w)$ the set of basis markings consistent with a given observation $w\in {\cal L}(G)$.



\begin{figure}
  \centering
  \includegraphics[width=0.35\textwidth]{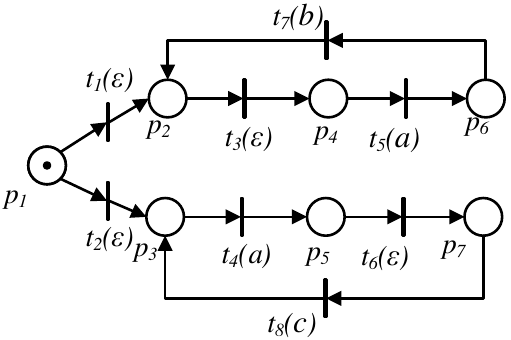}\\
  \caption{The LPN system in Example~\ref{eg:miniExp}.}\label{fig:LPN1}
\end{figure}

\exm\label{eg:miniExp}
Let us consider the LPN system in Fig.~\ref{fig:LPN1}, where $T_o=\{t_4, t_5, t_7, t_8\}$, $T_u=\{t_1, t_2, t_3, t_6\}$. Transition $t_4$ and $t_5$ are labeled by $a$, transition $t_7$ is labeled by $b$, and transition $t_8$ is labeled by $c$.
At the initial marking $M_0=[1\ 0\ 0\ 0\ 0\ 0\ 0]^T$, the minimal explanations of $t_4$ is $\Sigma_{min}(M_0,t_4)=\{t_2\}$, and thus $Y_{min}(M_0,t_4)=\{[0\ 1\ 0\ 0]^T\}$. The corresponding basis marking is $M_0+C(\cdot,t_4)+C_u\cdot [0\ 0\ 1\ 0]^T=M_1=[0\ 0\ 0\ 0\ 1\ 0\ 0]^T$.
At $M_0$, the minimal explanation of $t_5$ is $\Sigma_{min}(M_0,t_5)=\{t_1t_3\}$, and thus $Y_{min}(M_0,t_5)=\{[1\ 0\ 1\ 0]^T\}$. The basis marking obtained is $M_0+C(\cdot,t_5)+C_u\cdot [1\ 0\ 1\ 0]^T=M_2= [0\ 0\ 0\ 0\ 0\ 1\ 0]^T$.
\hfill $\diamond$
\eexm

\prop\label{prop:basis}\cite{cabasino2011discrete}
Let $G=(N,M_0,E,\ell)$ be an LPN whose $T_u$-induced subnet is acyclic, $M_b\in {\cal M}_b$ a basis marking of $G$, and $w\in {\cal L}(G)$ an observation generated by $G$. We have
\begin{enumerate}
  \item a marking $M$ is reachable from $M_b$ if and only if
\begin{equation}\label{eq:basis}
M=M_b+C_u\cdot y_u
\end{equation}
has a nonnegative solution $y_u\in \mathbb{N}^{n_u}$.
  \item ${\cal C}(w)=\bigcup_{M_b\in {\cal C}_b(w)} \{M\in \mathbb{N}^m|M=M_b+C_u\cdot y_u:y_u\in  \mathbb{N}^{n_u}\}$
\end{enumerate}
\eprop

Statement 1) of Proposition~\ref{prop:basis} implies that any solution $y_u \in \mathbb{N}^{n_u}$ of Eq.~\eqref{eq:basis} corresponds to the firing vector of a firable sequence $\sigma$ from $M_b$, i.e., $M_b[\sigma\rangle$ and $\pi(\sigma)=y_u$. According to Statement 2), the set of markings consistent with an observation can be characterized using
linear algebra without an exhaustive marking enumeration.

\subsection{Detectability}

In this subsection we recall the definitions of the detectability problems of the LPN. As in \cite{tong2019verification}, we make the following two assumptions: 1) the LPN system $G$ is deadlock free. 2) the $T_u$-induced subnet of $G$ is acyclic.
For more details, we refer to \cite{tong2019verification}.

\dfn\label{def:SD}
[Strong detectability]
An LPN system $G=(N,M_0,\allowbreak E,\ell)$ is said to be \emph{strongly detectable}
if
$$\exists K\in \mathbb{N},\forall \sigma\in L^\omega(G), \forall \sigma'\preceq \sigma, |w|\geq K \Rightarrow |{\cal C}(w)|=1,$$
where $w=\ell(\sigma')$. \hfill $\diamond$
\edfn

An LPN system is strongly detectable if the current and the subsequent states of the system can be determined after a finite number of events observed for all trajectories of the system.

\dfn\label{def:WD}
[Weak detectability]
An LPN system $G=(N,M_0,\allowbreak E,\ell)$ is said to be \emph{weakly detectable}
if
$$ \exists K\in \mathbb{N},\exists \sigma\in  L^\omega(G), \forall \sigma'\preceq \sigma, |w|\geq K \Rightarrow |{\cal C}(w)|=1,$$
where $w=\ell(\sigma')$. \hfill $\diamond$
\edfn

An LPN system is weakly detectable if we can determine, after a finite number of observations,
the current and subsequent states of the system for some trajectories of the system.

\dfn\label{def:SPD}
[Periodically strong detectability]
An LPN system $G=(N,M_0,\allowbreak E,\ell)$ is said to be \emph{periodically strongly detectable} if
$\exists K\in \mathbb{N}, \forall \sigma\in L^\omega (G),\forall \sigma' \preceq \sigma$,
$$  \exists \sigma''\in T^*: \sigma'\sigma''\preceq \sigma \wedge |\ell(\sigma'')|<K \Rightarrow |{\cal C}(w)|=1,$$
where $w=\ell(\sigma'\sigma'')$. \hfill $\diamond$
\edfn

An LPN system is periodically strongly detectable if the current and the subsequent states of the system can be periodically determined for all trajectories of the system.

\dfn\label{def:WPD}
[Periodically weak detectability]
An LPN system $G=(N,M_0,\allowbreak E,\ell)$ is said to be \emph{periodically weakly detectable}
if
$\exists K\in \mathbb{N}$, $\exists \sigma\in L^\omega (G),\forall \sigma' \preceq \sigma$,
$$  \exists \sigma''\in T^*: \sigma'\sigma''\preceq \sigma \wedge |\ell(\sigma'')|<K \Rightarrow |{\cal C}(w)|=1,$$
where $w=\ell(\sigma'\sigma'')$. \hfill $\diamond$
\edfn

An LPN system is periodically weakly detectable if we can periodically determine the current state of the system for some trajectories of the system.

In the case of labeled bounded Petri net systems, the detectability verification algorithm presented in \cite{tong2019verification} relies on the construction of the observer from the basis reachability graph of the Petri net system, a step that requires exponential complexity in the worst case. Therefore, next we will show that our approach for detectability analysis based on the verifier net is more efficient than the previous approach of building an observer from the reachability graph of the Petri net system.

\section{Analysis by verifier nets}\label{sec:ver}

In this section, we introduce the construction algorithm of the verifier net as well as its property, and then we build the BRG of the verifier net.

The notion of verifier net was first proposed in \cite{cabasino2012new} for verification of diagnosability, and there are other literatures use the similar approach to test diagnosability \cite{madalinski2010diagnosability}, prognosability \cite{yin2018verification} and the detectability \cite{masopust2018deciding}. Our approach is similar to \cite{masopust2018deciding} and \cite{cabasino2012new}, however, the difference here is that our verifier nets is a special labeled Petri net, witch is not in \cite{masopust2018deciding}. And we modified the labeling function of the verifier net for detectability, that is, there is no fault transitions, and the domain and range of our verifier nets is different from \cite{cabasino2012new}.


\subsection{Verifier Net}

Let $G=(N,M_0,E,\ell)$ be an LPN system, where $N=(P, T, Pre, Post)$, $T=T_o \cup T_u$, $\ell:T\rightarrow E\cup\{\eps\}$.
We donate $G'=(N',M_0',E,\ell')$ be a place-disjoint copy of $G$, that is, the initial marking $M_0'=M_0$, the event set of $G'$ is identical to the alphabet $E$, the labeling function of $G'$ is equal to $\ell$ restricted to $T'$, and $N'=(P', T', Pre', Post')$ be its $T'$-induced subnet, where $T'=T$. To distinguish among places of $N$ and $N'$, we denote them as $P$ and $P'$, respectively.


\lem\label{lem:copy-G}
Let $G=(N,M_0,E,\ell)$ be an LPN system, $G'=(N',M_0',E,\ell')$ the place-disjoint copy of $G$. If there exists a sequence $\sigma'=t_{j1}'t_{j2}'\cdots t_{jk}' \in {T'}^*$ in $G'$, that $M_j'[\sigma' \rangle M_k'$ where $M_j',M_k'\in R(N',M_0')$, then there must exist a sequence $\sigma=t_{j1}t_{j2}\cdots t_{jk}\in T^*$ in $G$, that $M_j[\sigma \rangle M_k$, $M_j=M_j'$, $M_k=M_k'$, where $M_j,M_k\in R(N,M_0)$ in $G$, and $\ell(\sigma) = \ell'(\sigma')$.
\elem
\prof
This result follows directly from the construction of $G'$. Since $G'$ is a place-disjoint copy of $G$, and $M_0'=M_0$, $T'=T$, $M_j\in R(N,M_0)$ and $M_j'\in R(N',M_0')$, there must exist a sequence $\sigma_j$ in $G$ and $\sigma_j'$ in $G'$ that $\ell(\sigma_j) = \ell'(\sigma_j')$, and $M_j=M_j'$
where $M_0[\sigma_j \rangle M_j$, $M_0'[\sigma_j' \rangle M_j'$.
Since there exists $\sigma'$ in $G'$, that $M_j'[\sigma' \rangle M_k'$, thus $M_0'[\sigma_j'\sigma' \rangle M_k'$. therefore, there must exist a sequence $\sigma$ in $G$, that $M_0[\sigma_j \sigma \rangle M_k$, $M_k=M_k'$, and $\ell(\sigma_j \sigma) = \ell'(\sigma_j' \sigma')$. Thus, $\ell(\sigma) = \ell'(\sigma')$.
\eprof


The Verifier Net (denoted by VN hereafter) system is the labeled Petri net system obtained by composing, in a manner made precise below, $G$ with $G'$ assuming that the synchronization is performed on the observable transition labels. We denote the VN system as $V=(\tilde{N},\tilde{M_0},E,\tilde{\ell})$, where $\tilde{N}=(\tilde{P}, \tilde{T}, \tilde{P}re, \tilde{P}ost)$ is a special Petri net, $\tilde{M_0}=$
   $\left[\begin{array}{cc}
  M_0'\\
  M_0
\end{array}\right]$ is the initial marking of $V$, $E$ is the alphabet, $\tilde{\ell}:\tilde{T} \rightarrow E\cup\{\eps\}$ is the labeling function of $V$. In net $\tilde{N}$, $\tilde{P}=P'\cup P$ is a set of places of $V$, and according to the labeling function $\tilde{\ell}$, let $\lambda$ be the empty transition, the set of transitions can be partitioned into two disjoint sets $\tilde{T}=\tilde{T}_o \cup \tilde{T}_u$, where $\tilde{T}_o = \{(t',t)|t' \in T_o', t\in T_o, \ell'(t')=\ell(t)\in E \}$ is the set of observable transitions, and $\tilde{T}_u = (T_u' \times \{\lambda\})\cup (\{\lambda\} \times T_u)$ is the set of unobservable transitions, i.e., $\tilde{\ell}(\tilde{T}_o)\in E$ and $\tilde{\ell}(\tilde{T}_u)\in \{\varepsilon \}$.


The function $\tilde{P}re: \tilde{P} \times \tilde{T} \rightarrow \mathbb{N}$ and $\tilde{P}ost: \tilde{P} \times \tilde{T} \rightarrow \mathbb{N}$ are defined in the Algorithm~\ref{alo:Verifier}.

\begin{algorithm}
\caption{Construction of the Verifier Net}
\label{alo:Verifier}
  \begin{algorithmic}[1]
   \Require
    A bounded LPN system $G=(N,M_0,E,\ell)$, where $N=(P, T, Pre, Post)$, $T=T_o \cup T_u$, $\ell:T\rightarrow E\cup\{\eps\}$.
   \Ensure
    VN labeled system $V=(\tilde{N},\tilde{M_0},E,\tilde{\ell})$, where $\tilde{N}=(\tilde{P}, \tilde{T}, \tilde{P}re, \tilde{P}ost)$, and $\tilde{\ell}:\tilde{T}\rightarrow E\cup\{\eps\}$.
   \State $G'=(N',M_0',E,\ell')$ be the labeled Petri net system defined as discussed above.
   \State $\tilde{P}=P'\cup P$, $\tilde{M_0}=$
   $\left[\begin{array}{cc}
  M_0'\\
  M_0
\end{array}\right]$.

   \ForAll{transitions $t_u \in T_u$,}
    \State $\bullet$ Add a transition $\tilde{t} \in \tilde{T}$ denoted as $(\lambda, t_u)$, $\tilde{\ell}(\tilde{t})=\varepsilon$;
    \State $\bullet$ for all $p\in P'$, let $\tilde{P}re(p,\tilde{t})=\tilde{P}ost(p,\tilde{t})=0$;
    \State $\bullet$ for all $p\in P$, let $\tilde{P}re(p,\tilde{t})=Pre(p,t_u)$ and $\tilde{P}ost(p,\tilde{t})=Post(p,t_u)$;
   \EndFor

   \ForAll{transitions $t_u' \in T_u'$,}
    \State $\bullet$ Add a transition $\tilde{t} \in \tilde{T}$ denoted as $(t_u',\lambda)$, $\tilde{\ell}(\tilde{t})=\varepsilon$;
    \State $\bullet$ for all $p\in P'$, let $\tilde{P}re(p,\tilde{t})=Pre'(p,t_u')$ and $\tilde{P}ost(p,\tilde{t})=Post'(p,t_u')$
    \State $\bullet$ for all $p\in P$, let $\tilde{P}re(p,\tilde{t})=\tilde{P}ost(p,\tilde{t})=0$;
   \EndFor

   \ForAll{labels $e \in E$,}
    \State $\bullet$ For any pair $t_o'$, $t_o$ with $t_o' \in T_o'$ , $t_o \in T_o$, $\ell'(t_o')=\ell(t_o)=e$;
    \State $\bullet$ Add a transition $\tilde{t}\in \tilde{T}$ denoted as $(t_o',t_o)$, $\tilde{\ell}(\tilde{t})=e$;
    \State $\bullet$ for all $p\in P'$, let $\tilde{P}re(p,\tilde{t})=Pre'(p,t_o')$ and $\tilde{P}ost(p,\tilde{t})=Post'(p,t_o')$;
    \State $\bullet$ for all $p\in P$, let $\tilde{P}re(p,\tilde{t})=Pre(p,t_o)$ and $\tilde{P}ost(p,\tilde{t})=Post(p,t_o)$.
   \EndFor

  \end{algorithmic}
\end{algorithm}

The VN, constructed by Algorithm~\ref{alo:Verifier}, is a labeled Petri net system. The initial marking $\tilde{M_0}$ is the concatenation of the initial marking of $G$ and $G'$ (Step 2). All the unobservable transitions are indicated with a pair $\tilde{t}=(\lambda, t_u)$ (Step 3 to 7) or $\tilde{t}=(t_u',\lambda)$ (Steps 8-12), where $t_u \in T_u$ in $G$, $t_u' \in T_u'$ in $G'$, and $\tilde{\ell}(\tilde{t})=\varepsilon$. All the observable transitions are indicated as $\tilde{t} = (t_o', t_o)$, where $t_o' \in T_o'$ in $G'$, $t_o\in T_o$ in $G$, and $\tilde{\ell}(\tilde{t})= \ell(t_o')=\ell(t_o)$ (Steps 13-18).

As the VN is a Petri net, thus the basics of Petri net in Section~\ref{sec:pre} are also suitable for the VN.
Given a VN system $V=(\tilde{N},\tilde{M_0},E,\tilde{\ell})$, the incidence of matix of $V$ is $\tilde{C}=\tilde{P}ost-\tilde{P}re$. Let $\tilde{N}'=(\tilde{P}', \tilde{T}', \tilde{P}re', \tilde{P}ost')$ be the $\tilde{T}_u$-induced subnet of $\tilde{N}$, where $\tilde{T}_u$ is the set of unobservable transitions. The incidence matrix of the $\tilde{T}_u$-induced subnet is denoted by $\tilde{C}_u=\tilde{P}ost'-\tilde{P}re'$.

\begin{figure}
  \centering
  \includegraphics[width=0.45\textwidth]{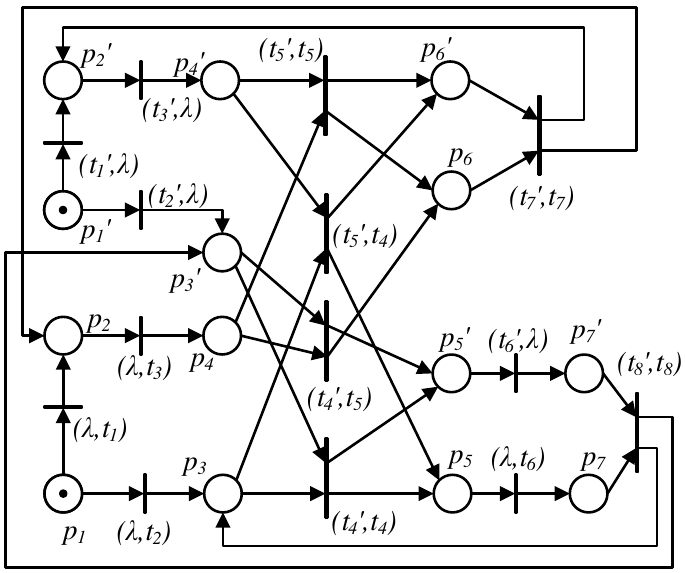}\\
  \caption{The VN system of the LPN in Fig.~\ref{fig:LPN1}.}\label{fig:VN1}
\end{figure}

\exm\label{eg:VN1}
Let us consider again the LPN system in Fig.~\ref{fig:LPN1}, by Algorithm~\ref{alo:Verifier}, its VN is presented in Fig.~\ref{fig:VN1}.
The set of places of the VN is obtained by the union of the set of places $P$ of the Petri net system $G$ in Fig.~\ref{fig:LPN1} and the set of places $P'$ of the $T'$-induced subnet. The $T'$-induced subnet is obtained from $G$. In VN, the initial marking $\tilde{M_0}= p_1'+p_1$, and there are eight unobservable transitions $\tilde{T}_u=\{(t_1',\lambda),(\lambda,t_1),(t_2',\lambda),(\lambda,t_2),(t_3',\lambda), (\lambda,t_3),\allowbreak (t_6',\lambda),\allowbreak(\lambda,t_6)\}$, and six observable transitions $\tilde{T}_o=\{(t_4',t_4),\allowbreak(t_5',t_5),\allowbreak(t_5',t_4),(t_4',t_5),(t_7',t_7), (t_8',t_8)\}$.
\hfill $\diamond$
\eexm

\lem\label{lem:seq-VN}
Let $G=(N,M_0,E,\ell)$ be an LPN system, $V=(\tilde{N},\tilde{M_0},E,\tilde{\ell})$ the VN of $G$. There exists a sequence $\tilde{\sigma}\in {\tilde{T}}^*$ in $V$, where $\tilde{\sigma}=(t_{j1}',t_{j1})(t_{j2}',t_{j2})\cdots (t_{jk}',t_{jk})$, if and only if there exists a sequence $\sigma'=t_{j1}'t_{j2}'\cdots t_{jk}'\in {T'}^*$ in $G'$, and a sequence $\sigma=t_{j1}t_{j2}\cdots t_{jk}\in T^*$ in $G$, and $\ell'(\sigma')=\ell(\sigma)$.
\elem
\prof
Follows from Algorithm~\ref{alo:Verifier}.
\eprof

In other words, VN is constructed by all pairs of sequences that have the same observation. Thus, for any sequence $\tilde{\sigma}=(\sigma',\sigma) \in L(\tilde{N},\tilde{M_0})$ in $V$, we can find $\sigma'$ in $G'$ and $\sigma$ in $G$, that they have the same observation $\ell'(\sigma')=\ell(\sigma)$. On the other hand, for any $\sigma' \in L(N',M_0')$ and $\sigma \in L(N,M_0)$, with $\ell'(\sigma')=\ell(\sigma)$, we can also find sequence
$\tilde{\sigma}=(\sigma',\sigma)$ in $V$.
\lem\label{lem:seq2-VN}
Let $G=(N,M_0,E,\ell)$ be an LPN system, $V=(\tilde{N},\tilde{M_0},E,\tilde{\ell})$ the VN of $G$. If there exists a sequence $\tilde{\sigma}\in L(\tilde{N},\tilde{M_0})$ in $V$, $\tilde{M_0}[\tilde{\sigma } \rangle \tilde{M} =$
$\left[\begin{array}{cc}
  M'\\
  M
\end{array}\right]$
that $M'\neq M$, then there must exist two different markings $M_1, M_2 \in R(N,M_0)$ in $G$ that $M_1 \neq M_2$, where $M_0[\sigma_1 \rangle M_1$,$M_0[\sigma_2 \rangle M_2$, $\ell(\sigma_1) = \ell(\sigma_2)$.
\elem
\prof
Since $\tilde{\sigma}\in L(\tilde{N},\tilde{M_0})$ in $V$, by Lemma~\ref{lem:seq-VN}, there must exist $\sigma' \in L(N',M_0')$ in $G'$ and $\sigma \in L(N,M_0)$ in $G$, with $\ell'(\sigma')= \ell(\sigma)$. $\tilde{M_0}[\tilde{\sigma } \rangle \tilde{M} =$
$\left[\begin{array}{cc}
  M'\\
  M
\end{array}\right]$,
according to the construction of VN, $M_0'[\sigma' \rangle M'$,$M_0[\sigma \rangle M$. Since $M_0'[\sigma' \rangle M'$ in $G'$, by Lemma~\ref{lem:copy-G}, there must exist a sequence $\sigma_1 \in L(N,M_0)$ in $G$ that $M_0[\sigma_1 \rangle M_1 = M'$, with $\ell(\sigma_1) = \ell'(\sigma')$. Since $\ell'(\sigma')= \ell(\sigma)$ and $M'\neq M$, therefore, $\ell(\sigma_1) = \ell'(\sigma')= \ell(\sigma)$ and $M_1 = M' \neq M$. Clearly, $\sigma_1 \neq \sigma$.
\eprof

In simple words, in an LPN system, if a marking $\tilde{M} \in R(\tilde{N},\tilde{M}_0)$ of its VN that $\tilde{M} =$
$\left[\begin{array}{cc}
  M'\\
  M
\end{array}\right]$
with $M'\neq M$, then there must exist two different sequences whose reachable markings are different from each other in the LPN system.

\lem\label{lem:seq3-VN}
Let $G=(N,M_0,E,\ell)$ be an LPN system, $V=(\tilde{N},\tilde{M_0},E,\tilde{\ell})$ the VN of $G$. If there exists two sequence $\tilde{\sigma}_1,\tilde{\sigma}_2\in L(\tilde{N},\tilde{M_0})$ in $V$, with $\tilde{\ell}(\tilde{\sigma}_1)=\tilde{\ell}(\tilde{\sigma}_2)$, $\tilde{M_0}[\tilde{\sigma }_1 \rangle \tilde{M_1}$, $\tilde{M_0}[\tilde{\sigma}_2 \rangle \tilde{M_2}$, that $\tilde{M_1} \neq \tilde{M_2}$, then there must exist two different markings $M_1, M_2 \in R(N,M_0)$ in $G$ that $M_1 \neq M_2$, where $M_0[\sigma_1 \rangle M_1$,$M_0[\sigma_2 \rangle M_2$, $\ell(\sigma_1) = \ell(\sigma_2)$.
\elem
\prof
Let $\tilde{M_1}=$
$\left[\begin{array}{cc}
  M_{r1}'\\
  M_{r1}
\end{array}\right]$,
$\tilde{M_2}=$
$\left[\begin{array}{cc}
  M_{r2}'\\
  M_{r2}
\end{array}\right]$,
where $M_0'[\sigma_{r1}' \rangle M_{r1}'$, $M_0[\sigma_{r1} \rangle M_{r1}$, $M_0'[\sigma_{r2}' \rangle M_{r2}'$, $M_0[\sigma_{r2} \rangle M_{r2}$. Thus $\ell(\sigma_{r1}') = \ell(\sigma_{r1})= \ell(\sigma_{r2}')= \ell(\sigma_{r2})$.
Since $\tilde{M_1}\neq \tilde{M}_2$, thus $M_{r1}',M_{r1},M_{r2}',M_{r2}$ can not all be equal. Thus there is at least one marking in $\{ M_{r1}',M_{r2}',M_{r2} \}$, that does not equal $M_{r1}$. There are three cases.

$\mathbf{Case 1}$: $M_{r1}' \neq M_{r1}$

Since $\tilde{M_1}=$
$\left[\begin{array}{cc}
  M_{r1}'\\
  M_{r1}
\end{array}\right]$, that $M_{r1}' \neq M_{r1}$, by Lemma~\ref{lem:seq2-VN}, there must exist two different markings $M_1, M_2 \in R(N,M_0)$ in $G$ that $M_1 \neq M_2$, where $M_0[\sigma_1 \rangle M_1$,$M_0[\sigma_2 \rangle M_2$, $\ell(\sigma_1) = \ell(\sigma_2)$.

$\mathbf{Case 2}$: $M_{r2}' \neq M_{r1}$

Since $\ell(\sigma_{r2}')= \ell(\sigma_{r1})$ ,according to the construction of $V$, there exists a marking $\tilde{M_3}=$
$\left[\begin{array}{cc}
  M_{r2}'\\
  M_{r1}
\end{array}\right]$, with $M_{r2}' \neq M_{r1}$. It is the same as $\mathbf{Case 1}$.

$\mathbf{Case 3}$: $M_{r2} \neq M_{r1}$

Since $\ell(\sigma_{r2})= \ell(\sigma_{r1})$ and $M_{r2} \neq M_{r1}$, that $M_{r2},M_{r1}$ is exactly the two different marking in $G$.
\eprof

In simple words, in an LPN system, if there exists two different marking with the same observation in its VN, then there must exist two different markings in the LPN system.

\prop\label{prop:suf}
Let $G=(N,M_0,E,\ell)$ be an LPN system, $V=(\tilde{N},\tilde{M_0},E,\tilde{\ell})$ the VN of $G$. The $\tilde{T}_u$-induced subnet of $V$ is acyclic, if and only if the $T_u$-induced subnet of $G$ is acyclic.
\eprop
\prof
(If) Assume the The $\tilde{T}_u$-induced subnet of $V$ is not acyclic. Clearly, there exists a cycle in the $\tilde{T}_u$-induced subnet, ie., there exists a sequence $\tilde{\sigma_u}=(t_{j1}',t_{j1})(t_{j2}',t_{j2})\cdots (t_{jk}',t_{jk}) \in \tilde{T_u}^*$, such that $\tilde{M}[\tilde{\sigma_u} \rangle \tilde{M}$. By Lemma~\ref{lem:seq-VN}, there must exist a sequence $\sigma=t_{j1}t_{j2}\cdots t_{jk} $ in $G$, $\ell(\sigma)= \tilde{\ell}(\tilde{\sigma})=\varepsilon$. Let $\tilde{M} =$
$\left[\begin{array}{cc}
  M'\\
  M
\end{array}\right]$,
thus $M[\sigma \rangle M$. Therefore, the $T_u$-induced subnet of $G$ is also not acyclic.

(Only if) Assume the $T_u$-induced subnet of $G$ is not acyclic. Clearly, there exists a cycle in the $T_u$-induced subnet, ie., there exists a sequence $\sigma_u=t_{j1}t_{j2}\cdots t_{jk} \in T_u^*$, such that $M[\sigma_u \rangle M$. By Lemma~\ref{lem:copy-G}, there also exists a sequence $\sigma'=t_{j1}'t_{j2}'\cdots t_{jk}'$ in $G'$ that $M'[\sigma' \rangle M'$, with $\ell'(\sigma')= \ell(\sigma_u)=\varepsilon$. By Lemma~\ref{lem:seq-VN}, there must exist a sequence $\tilde{\sigma}=(t_{j1}',t_{j1})(t_{j2}',t_{j2})\cdots (t_{jk}',t_{jk}) $ and a marking $\tilde{M} =$
$\left[\begin{array}{cc}
  M'\\
  M
\end{array}\right]$,
such that $\tilde{M}[\tilde{\sigma } \rangle \tilde{M}$ with $\tilde{\ell}(\tilde{\sigma})=\ell'(\sigma')=\ell(\sigma_u)= \varepsilon$. Therefore, The $\tilde{T}_u$-induced subnet of $V$ is not acyclic.
\eprof

\textbf{Remark}: In the next subsection, we need to construct the BRG of the VN. To use the construction approach in \cite{cabasino2011discrete}, the $T_u$-induced subnet of the Petri net need to be acyclic. By Proposition~\ref{prop:suf}, to construct the BRG of VN, we just need to insure the $T_u$-induced subnet of the LPN system is acyclic.

\subsection{Construction of the BRG}

In the framework of VN, the \emph{reachability graph} (RG) of VN is usually constructed to verify its property, e.g., the diagnosability \cite{madalinski2010diagnosability,cabasino2012new}, prognosability \cite{yin2018verification} and the detectability \cite{masopust2018deciding}. It is known that, the complexity of constructing the RG of a Petri net system is exponential to its size\footnote{The size of a Petri net system usually refers the number of places, the number of transitions, and the number of initial tokens, etc.}.
Therefore, to verify detectability of large-scaled systems the state explosion problem cannot be avoided. In this subsection, we construct BRG to verifying detectability without enumerating all states of the system. As illustrated in \cite{tong2017verification,ziyue2017basis}, the BRG of a Petri net is usually much smaller than its corresponding RG.
In this way, the state explosion problem is practically avoided \cite{tong2015verification}.

Based on the notion of basis markings, we introduce the \emph{basis reachability graph} (BRG) for VN. To guarantee that the BRG is finite, we assume that the LPN system is bounded.
Let $\tilde{M_b}\in \tilde{{\cal M}_b}$ be the basis marking of VN, for each basis marking $\tilde{M_b}$, one value is assigned by functions $\alpha:\tilde{{\cal M}_b}\rightarrow \{0,1\}$ that is defined by
Eq.~\eqref{eq:alpha}.

\begin{equation}\label{eq:alpha}
\alpha(\tilde{M_b})=\left\{\begin{array}{ll}
  1 & \text{if $\tilde{M_b} + \tilde{C}_u \cdot y_u \geq 0$ has a positive }\\
    & \text{integer solution;}\\
  0 & \text{otherwise.}
\end{array}\right.
\end{equation}

We denote $B = (X, E, f, x_0)$ the BRG for a VN system $V=(\tilde{N},\tilde{M_0},E,\tilde{\ell})$, where $X\subseteq \tilde{{\cal M}}_b\times \{0,1\}$ is a finite set of states, and each state $x\in X$ of the BRG is a pair $(\tilde{M_b},\alpha(\tilde{M_b}))$. We denote the $i$-th (with $i=1,2$) element of $x$ as $x(i)$. The initial node of the BRG is $x_0=(\tilde{M_0},\alpha(\tilde{M_0}))$. The event set of the BRG is identical to the alphabet $E$. The transition function $f: X\times E \rightarrow X$ can be determined by the following rule. If at marking $\tilde{M_{b}}$ there is an observable transition $t$ for which a minimal explanation exists and the firing of $t$ and one of its minimal explanations lead to $\tilde{M_{b}'}$, then an edge from node $(\tilde{M_{b}},\alpha(\tilde{M_{b}}))$ to node $(\tilde{M_{b}'},\alpha(\tilde{M_{b}'}))$ labeled with $\ell(t)$ is defined in the BRG.
By construction of the VN, we know that the VN is a special labeled Petri net, thus, the BRG of the VN can be constructed by applying the algorithm in \cite{tong2019verification}.

%

\begin{figure}
  \centering
  \includegraphics[width=0.35\textwidth]{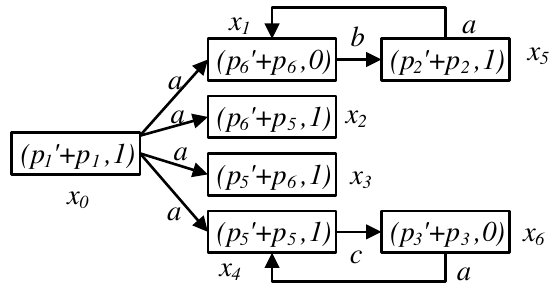}\\
  \caption{The BRG of the VN in Fig.~\ref{fig:VN1}.}\label{fig:BRG1}
\end{figure}

\exm\label{eg:BRG1}
Consider again the LPN system in Fig.~\ref{fig:LPN1}, Where its VN in Fig.~\ref{fig:VN1} is already introduced in Example~\ref{eg:VN1}. The VN system has $25$ reachable markings while $7$ basis markings.
For basis marking $\tilde{M_0}=p_1'+p_1$, the Eq.~\eqref{eq:alpha} has $15$ solutions. In this case, $\alpha(\tilde{M_0})=1$.
For basis marking $\tilde{M_1}=p_6'+p_6$, by Eq.~\eqref{eq:alpha}, the equation does not have a positive integer solution. Therefore, $\alpha(\tilde{M_1})=0$.
The builded BRG of the VN is presented in Fig.~\ref{fig:BRG1}.
\hfill $\diamond$
\eexm

\lem\label{lem:brg1}
Let $G=(N,M_0,E,\ell)$ be an LPN system whose $T_u$-induced subnet is acyclic, $V=(\tilde{N},\tilde{M_0},E,\tilde{\ell})$ the VN of $G$, and $\tilde{M_b}\in \tilde{{\cal M}_{b}}$ a basis marking of $V$. If $\alpha(\tilde{M_b})=1$, then there exists an observation $w\in E^*$ of $G$ such that $|{\cal C}(w)|\neq 1$.
\elem
\prof
By assumption $\alpha(\tilde{M_b})=1$, Eq.~\eqref{eq:alpha} has a positive integer solution, thus there is a marking $\tilde{M}$ reachable from $\tilde{M}_b$ by firing unobservable transitions $\tilde{\sigma}_u\in \tilde{T^*_u}$ whose corresponding firing vector is $y_u=\pi(\tilde{\sigma}_u)$. Since the $T_u$-induced subnet of $G$ is acyclic, by proposition~\ref{prop:suf}, the $\tilde{T}_u$-induced subnet of $V$ is acyclic.
Let $\tilde{\sigma}\in \tilde{T^*}$ be transition sequence such that $\tilde{M_0}[\tilde{\sigma}\rangle \tilde{M}_b$ and $\tilde{\ell}(\tilde{\sigma})=w$. Clearly, $\tilde{M_0}[\tilde{\sigma}\tilde{\sigma}_u\rangle \tilde{M}$ and $\tilde{\ell}(\tilde{\sigma} \tilde{\sigma}_u)=\tilde{\ell}(\tilde{\sigma})=w$. Since the $\tilde{T}_u$-induced subnet of $V$ is acyclic and $y_u\neq \vec{0}$, thus $\tilde{M}\neq \tilde{M}_b$.
By Lemma~\ref{lem:seq3-VN}, there must exist two different markings $M_1, M_2 \in R(N,M_0)$ in $G$ that $M_1 \neq M_2$, where $M_0[\sigma_1 \rangle M_1$,$M_0[\sigma_2 \rangle M_2$, $\ell(\sigma_1) = \ell(\sigma_2)=\tilde{\ell}(\tilde{\sigma})=w$. Therefore, $M_1,M_2 \in {\cal C}(w)$, thus $|{\cal C}(w)|\neq 1$.
\eprof

In simple words, if $\alpha(\tilde{M_b})=1$, there is an observation $w$ such that ${\cal C}(w)$ contains more than one marking.


\lem\label{lem:brg2}
Let $G=(N,M_0,E,\ell)$ be an LPN system, $V=(\tilde{N},\tilde{M_0},E,\tilde{\ell})$ the VN of $G$, and $\tilde{M_b}\in \tilde{{\cal M}_{b}}$ a basis marking of $V$. If $\tilde{M_b}=$
$\left[\begin{array}{cc}
  M_b'\\
  M_b
\end{array}\right]$
$\in R(\tilde{N},\tilde{M_0})$
with $M_b' \neq M_b$, then there exists an observation $w\in E^*$ of $G$ such that $|{\cal C}(w)|\neq 1$.
\elem
\prof
Since $\tilde{M_b}=$
$\left[\begin{array}{cc}
  M_b'\\
  M_b
\end{array}\right]$
$\in R(\tilde{N},\tilde{M_0})$
with $M_b' \neq M_b$, by Lemma~\ref{lem:seq2-VN}, there must exist two different markings $M_1, M_2 \in R(N,M_0)$ in $G$ with $M_1 \neq M_2$, where $M_0[\sigma_1 \rangle M_1$,$M_0[\sigma_2 \rangle M_2$, $\ell(\sigma_1) = \ell(\sigma_2)$. Let the observation $w= \ell(\sigma_1) = \ell(\sigma_2)$, where $w\in E^*$ of $G$. Therefore, $M_1,M_2 \in {\cal C}(w)$, thus $|{\cal C}(w)|\neq 1$.
\eprof

In simple words, if $\tilde{M_b}=$
$\left[\begin{array}{cc}
  M_b'\\
  M_b
\end{array}\right]$
with $M_b' \neq M_b$, there is an observation $w$ such that ${\cal C}(w)$ contains more than one marking.

\prop\label{prop:brg}
Let $G=(N,M_0,E,\ell)$ be an LPN system, $V=(\tilde{N},\tilde{M_0},E,\tilde{\ell})$ the VN of $G$, there exists an observation $w\in E^*$ of $G$ that $|{\cal C}(w)|\neq 1$, if and only if there exists a basis marking $\tilde{M_b}\in \tilde{{\cal M}_{b}}$ of $V$, such that $\alpha(\tilde{M_b})=1$ or $\tilde{M_b}=$
$\left[\begin{array}{cc}
  M_b'\\
  M_b
\end{array}\right]$
that $M_b' \neq M_b$.
\eprop

\prof
(If)
Follows from Lemma~\ref{lem:brg1} and \ref{lem:brg2}.

(Only if)
Assume that there exists an observation $w\in E^*$ of $G$ that $|{\cal C}(w)|\neq 1$, there exists two different markings $M_1, M_2 \in {\cal C}(w)$ in $G$ with $M_1 \neq M_2$, where $M_0[\sigma_1 \rangle M_1$,$M_0[\sigma_2 \rangle M_2$, $\ell(\sigma_1) = \ell(\sigma_2) = w$.
According to the construction of the VN, there must exist a sequence $\tilde{\sigma}\in L(\tilde{N},\tilde{M_0})$ in $V$, $\tilde{M_0}[\tilde{\sigma } \rangle \tilde{M} =$
$\left[\begin{array}{cc}
  M'\\
  M
\end{array}\right]$
that $M'\neq M$, Since $\tilde{M_0}[\tilde{\sigma } \rangle \tilde{M}$, $\tilde{M} \in R(\tilde{N},\tilde{M_0})$, by the construction of BRG, therefore, $\alpha(\tilde{M_b})=1$ or $\tilde{M_b}=$
$\left[\begin{array}{cc}
  M_b'\\
  M_b
\end{array}\right]$
that $M_b' \neq M_b$.
\eprof

In simple words, in an LPN system, there exists an observation $w$ such that ${\cal C}(w)$ contains more than one marking, if and only if there exists a basis marking $\tilde{M_b}$ in the BRG of its VN, that either $\alpha(\tilde{M_b})=1$ or $\tilde{M_b}=$
$\left[\begin{array}{cc}
  M_b'\\
  M_b
\end{array}\right]$
with $M_b' \neq M_b$.

\coro\label{prop:all}
Let $G=(N,M_0,E,\ell)$ be an LPN system, $V=(\tilde{N},\tilde{M_0},E,\tilde{\ell})$ the VN of $G$.
If for all basis markings $\tilde{M_b} \in \tilde{{\cal M}_b}$ of $V$, that $\alpha(\tilde{M_{b}})=0$ and $\tilde{M_b}=$
$\left[\begin{array}{cc}
  M_b'\\
  M_b
\end{array}\right]$
with $M_b' = M_b$, then the LPN system is strongly detectable, weakly detectable, periodically strongly detectable, and periodically weakly detectable.
\ecoro

\prop\label{prop:suf}
Let $G=(N,M_0,E,\ell)$ be an LPN system, $V=(\tilde{N},\tilde{M_0},E,\tilde{\ell})$ the VN of $G$. System $G$ does not perform any detectability if for all basis markings $\tilde{M_b} \in \tilde{{\cal M}_b}$ of $V$, $\alpha(\tilde{M_b})=1$.
\eprop
\prof
Since $\forall \tilde{M_b} \in \tilde{{\cal M}_b}$ of $V$, $\alpha(\tilde{M_b})=1$, by Lemma~\ref{lem:brg1}, for all observations $w\in {\cal L}(G)$, ${\cal C}(w)$ is not a singleton, i.e., $|{\cal C}(w)| \neq 1$. Therefore, it is not possible for the system to have any detectability defined in Definitions~\ref{def:SD}, \ref{def:WD}, \ref{def:SPD} and \ref{def:WPD}.
\eprof

\section{Verification of detectability}\label{sec:verfication}

In this section, we show how the detectability of an bounded Petri net system can be efficiently checked by analyzing the BRG of the verifier net.

Since detectability considers the transition sequences of infinite length, and the BRG has a finite number of nodes, thus these transition sequences must contain cycles. Therefore, to present the necessary and sufficient conditions for detectability, we first study the properties of cycles in the BRG.

\dfn\label{def:cycle}[Simple cycle]
A \emph{(simple) cycle} in the BRG $B = (X, E, f, x_0)$ of a VN is a path $\gamma_j=x_{j1}e_{j1}x_{j2}\ldots x_{jk}\allowbreak e_{jk}x_{j1}$ that starts and ends at the same state but without repeat edges, where $x_{ji}\in X$ and $e_{ji}\in E$ with $i=\{1,2,\ldots,k\}$, and $\forall m,n\in\{1,2,\ldots,k\}, x_{jm}\neq x_{jn}$ where $m \neq n$. The corresponding observation of the cycle is
$w=e_{j1}\ldots e_{jk}$. A state $x_{ji}$ contained in $\gamma_j$ is denoted by $x_{ji}\in \gamma_j$. The set of simple cycles in the BRG is denoted by $\Gamma$. \hfill $\diamond$
\edfn

\subsection{Strong detectability}\label{subsec:SD}

\them\label{therm:S-detect}
Let $G=(N,M_0,E,\ell)$ be an LPN system whose $T_u$-induced subnet is acyclic, $V=(\tilde{N},\tilde{M_0},E,\tilde{\ell})$ the VN of $G$, and $B = (X, E, f, x_0)$ the BRG of $V$. LPN $G$ is strongly detectable if and only if for any $x=(\tilde{M_{b}},\alpha(\tilde{M_{b}})) \in X$ reachable from any cycle in $B$, that $\alpha(\tilde{M_{b}})=0$ and $\tilde{M_b}=$
$\left[\begin{array}{cc}
  M_b'\\
  M_b
\end{array}\right]$
with $M_b' = M_b$.
\ethem

\prof
(If) Assume LPN system $G$ is not strongly detectable, that is $\forall K\in \mathbb{N}, \exists \sigma\in L^\omega(G)$, $\exists \sigma'\preceq \sigma$, $w'=\ell(\sigma'), |w'|\geq K, \Rightarrow |{\cal C}(w')|\neq1$.
$\exists \sigma\in L^\omega(G)$ and $\exists \sigma'\preceq \sigma$, by Lemma~\ref{lem:seq-VN}, there must exist a sequence $\tilde{\sigma}\in L^\omega(V)$ and exist $\tilde{\sigma'}\preceq \tilde{\sigma}$, with $\tilde{\ell}(\tilde{\sigma})=\ell(\sigma)$, $\tilde{\ell}(\tilde{\sigma'})=\ell(\sigma')=w'$.
Since $\tilde{\sigma}$ is of an infinite length and $B$ has a finite number of nodes, the path along $\tilde{\ell}(\tilde{\sigma})=w$ must contain a cycle $\gamma_j=x_{j1}e_{j1}x_{j2}\ldots x_{jk}\allowbreak e_{jk}x_{j1}\in \Gamma$, i.e., there exist $w_1,w_2\in E^*$ such that $w=w_1(e_{j1}e_{j2}\ldots e_{jk})^*w_2$ and $|w_1|$ is finite.
Since the $T_u$-induced subnet of $G$ is acyclic, by proposition~\ref{prop:suf}, the $\tilde{T}_u$-induced subnet of $V$ is acyclic.
Thus let $\tilde{\ell}(\tilde{\sigma'})=w_1(e_{j1}e_{j2}\ldots e_{jk})^*(e_{j1}e_{j2}\ldots e_{jr})w''$, $|\tilde{\ell}(\tilde{\sigma'})|\geq K$($\forall K\in \mathbb{N}$), where $w''\preceq (e_{jr+1} \ldots e_{jk})(e_{j1}e_{j2}\ldots e_{jk})^*w_2$, and states $x_{jr}=f({\cal M}_b,w_1(e_{j1}e_{j2}\ldots e_{jk})^*(e_{j1}e_{j2}\ldots e_{jr}))$. Let $w_0=w_1(e_{j1}e_{j2}\ldots e_{jk})^*(e_{j1}e_{j2}\ldots e_{jr})$, thus $x_{jr}=f(x_0,w_0)$, $\ell(\sigma')=w_0w''$. Therefore, $f(x_{jr},w'')=f(x_0,w_0w'')=f(x_0,w')= (\tilde{M_{b}},\alpha(\tilde{M_{b}}))$.
Since $|{\cal C}(w')|\neq1$, $w'=\ell(\sigma')=w_0w''$, by Proposition~\ref{prop:brg}, $\alpha(\tilde{M_b})=1$ or $\tilde{M_b}=$
$\left[\begin{array}{cc}
  M_b'\\
  M_b
\end{array}\right]$
that $M_b' \neq M_b$.


(Only if)
Assume there exists $\gamma_j\in \Gamma$, $\exists x_{jr}\in \gamma_j$, $\exists w'\in E^*$ that $f(x_{jr},w')$ is defined, either $f(x_{jr},w')=(\tilde{M_{b}},1)$ or $\tilde{M_b}=$
$\left[\begin{array}{cc}
  M_b'\\
  M_b
\end{array}\right]$
with $M_b' \neq M_b$.
Clearly, there exists $\tilde{\sigma}\in L^\omega(V)$ and $w=\tilde{\ell}(\tilde{\sigma})$, $w_1,w_2\in E^*$ such that $w=w_1(e_{j1}e_{j2}\ldots e_{jk})^*w_2$ and $|w_1|$ is finite.
Since the $T_u$-induced subnet of $G$ is acyclic, by proposition~\ref{prop:suf}, the $\tilde{T}_u$-induced subnet of $V$ is acyclic.
Thus there exist $\tilde{\sigma'} \preceq \tilde{\sigma}$ with $\tilde{\ell}(\tilde{\sigma'})=w_1(e_{j1}e_{j2}\ldots e_{jk})^*(e_{j1}e_{j2}\ldots e_{jr})w'$, $|\tilde{\ell}(\tilde{\sigma'})|\geq K$($\forall K\in \mathbb{N}$), where $w'\preceq (e_{jr+1} \ldots e_{jk})(e_{j1}e_{j2}\ldots e_{jk})^*w_2$, and states in $f(x_0,w_1(e_{j1}e_{j2}\ldots e_{jk})^*(e_{j1}e_{j2}\ldots e_{jr}))=x_{jr}$. Let $w_0=w_1(e_{j1}e_{j2}\ldots e_{jk})^*(e_{j1}e_{j2}\ldots e_{jr})$, thus $f(M_0,w_0)=x_{jr}$.
Since the sequence $\tilde{\sigma}\in L^\omega(V)$ and $\tilde{\sigma'}\preceq \tilde{\sigma}$, by Lemma~\ref{lem:seq-VN}, there must exist a sequence $\sigma\in L^\omega(G)$ and exist $\sigma'\preceq \sigma$, with $\ell(\sigma')=\tilde{\ell}(\tilde{\sigma'})=w'$.
By assumption, we have $w'\in E^*$ that $f(x_{jr},w')$ is defined, $f(x_{jr},w')=(\tilde{M_{b}},1)$ (ie., $\alpha(\tilde{M_b})=1$) or $\tilde{M_b}=$
$\left[\begin{array}{cc}
  M_b'\\
  M_b
\end{array}\right]$
with $M_b' \neq M_b$.
Sice $f(x_0,w_0w')=f(x_{jr},w')$, by Proposition~\ref{prop:brg}, therefore $|{\cal C}(w_0w')|\neq1$.
\eprof

In words, an LPN system is strongly detectable if and only if in the BRG of its VN, all states reachable from a cycle have the form $\left(\left[\begin{array}{cc}
  M_b'\\
  M_b
\end{array}\right],0\right)$
with $M_b' = M_b$.

\exm\label{eg:SD}
Consider again the LPN system in Fig.~\ref{fig:LPN1}. Its VN is shown in Fig.~\ref{fig:VN1}, and the BRG of the VN is shown in Fig.~\ref{fig:BRG1}. Now we use Theorem~\ref{therm:S-detect} to check its strong detectability. In the BRG, we can see that $x_5(2)=1$ and $x_4(2)=1$, thus there are no cycles having all states $(\tilde{M_{b}},\alpha(\tilde{M_{b}}))$ that $\alpha(\tilde{M_{b}})=0$ and $\tilde{M_b}=$
$\left[\begin{array}{cc}
  M_b'\\
  M_b
\end{array}\right]$
with $M_b' = M_b$.
Therefore, the LPN system is not strongly detectable.
\hfill $\diamond$
\eexm

\subsection{Periodically strong detectability}\label{subsec:PSD}

An LPN system is said to be periodically strongly detectable if the current and the subsequent states of the system can be periodically determined for all trajectories of the system. Thus, to verify the periodically strong detectability, we need to check all the cycles in the BRG.

We define the set of states in a cycle of the BRG, that are not confused with the other states which is connected with the cycle, as marked states $X_m$. The marked states are constructed as Algorithm~\ref{alo:xm}.

\begin{algorithm}
\caption{Computation of the $X_m$ for BRG}
\label{alo:xm}
  \begin{algorithmic}[1]
   \Require
    A BRG $B = (X, E, f, x_0)$.
   \Ensure
    The corresponding marked states $X_m$.
   \State Let $\Gamma$ be the set of simple cycles in $B$.
   \ForAll{cycles $\gamma_j \in \Gamma$,}
     \ForAll{states $x_{jr} \in \gamma_j$,}
       \State Let $x_{jr} =(\tilde{M_{b}},\alpha(\tilde{M_{b}}))$ with $\tilde{M_b}=$
              $\left[\begin{array}{cc}
                M_b'\\
                M_b
              \end{array}\right]$.
       \If {$\alpha(\tilde{M_{b}})=0 \&\& M_b' = M_b$,}
         \State Let $x_{jr+1} \in \gamma_j$ be the next states of $x_{jr}$;
         \State Let $w \in E^*$ in $\gamma_j$ that $f(x_{jr+1},w)=x_{jr}$;
           \If {$\forall w' \in E^*$ in $B$, with $w'=w$, that $f(x_{jr+1},w')=x_{jr}$}
             \State Add state $x_{jr} \in X_m$
           \EndIf
       \EndIf
     \EndFor
   \EndFor
  \end{algorithmic}
\end{algorithm}

\prop\label{prop:xm}
Let $G=(N,M_0,E,\ell)$ be an LPN system whose $T_u$-induced subnet is acyclic, $V=(\tilde{N},\tilde{M_0},E,\tilde{\ell})$ the VN of $G$, $B = (X, E, f, x_0)$ the BRG of $V$, and $\gamma_j \in \Gamma$ a cycle in $B$.
Given a state $x_{jr}\in \gamma_j$, if $x_{jr} \notin {X}_m$, there exists an observation $w$ such that $|{\cal C}(w)|\neq 1$.
\eprop

\prof
According to the construction of $X_m$, we assume the state $x_{jr} =(\tilde{M_{b}}, \alpha(\tilde{M_{b}}))$ with $\tilde{M_b}=$
$\left[\begin{array}{cc}
 M_b'\\
 M_b
\end{array}\right]$, and let $x_{jr+1} \in \gamma_j$ be the next states of $x_{jr}$, let $w_1 \in E^*$ in $\gamma_j$ that $f(x_{jr+1},w_1)=x_{jr}$.
Since $x_{jr} \notin {X}_m$, by Algorithm~\ref{alo:xm}, $\alpha(\tilde{M_{b}})=1$ or $M_b' \neq M_b$ (Case 1) or there exists $w_1' \in E^*$ in $B$, with $w_1'=w_1$, that $f(x_{jr+1},w_1') \neq x_{jr}$ (Case 2). For these two cases, we prove that there exists an observation $w$ such that $|{\cal C}(w)|\neq 1$.

\textbf{Case 1}: $\alpha(\tilde{M_{b}})=1$ or $M_b' \neq M_b$

Since state $x_{jr} =(\tilde{M_{b}}, \alpha(\tilde{M_{b}}))$ with $\tilde{M_b}=$
$\left[\begin{array}{cc}
 M_b'\\
 M_b
\end{array}\right]$, such that $\alpha(\tilde{M_b})=1$ or $M_b' \neq M_b$. By Proposition~\ref{prop:brg}, therefore, there exists an observation $w\in E^*$ of $G$ such that $|{\cal C}(w)|\neq 1$.

\textbf{Case 2}: There exists $w_1' \in E^*$ in $B$, with $w_1'=w_1$, that $f(x_{jr+1},w_1') \neq x_{jr}$

Let $w_0' \in E^*$ in $B$, that $f(x_0,w_0')=x_{jr+1}$. Since $f(x_{jr+1},w_1)=x_{jr}$, thus $f(x_0,w_0'w_1)=f(x_{jr+1},w_1)=x_{jr}$. By assumption, $f(x_{jr+1},w_1') \neq x_{jr}$ with $w_1'=w_1$, let $x_i=f(x_{jr+1},w_1')$, thus $f(x_0,w_0'w_1')=f(x_{jr+1},w_1')=x_i \neq x_{jr}$.
Let $x_i=(\tilde{M_{b}'}, \alpha(\tilde{M_{b}}'))$, thus $\tilde{M_{b}'} \neq \tilde{M_{b}}$.
By Lemma~\ref{lem:seq3-VN}, there must exist two different markings $M_1, M_2 \in R(N,M_0)$ in $G$ that $M_1 \neq M_2$, where $M_0[\sigma_1 \rangle M_1$,$M_0[\sigma_2 \rangle M_2$, $\ell(\sigma_1) = \ell(\sigma_2)$. Let $w=\ell(\sigma_1) = \ell(\sigma_2)$, therefore, $M_1,M_2 \in {\cal C}(w)$, thus $|{\cal C}(w)|\neq 1$.
\eprof

In simple words, if a state, in the cycle of the BRG, is not a marked state, there is an observation $w$ such that ${\cal C}(w)$ contains more than one marking.
However, if the state $x_{jr}$ is a marked state, it does not mean there is an observation $w$ such that ${\cal C}(w)$ contains only one marking. Because in the BRG there may exist another state, which is not connected with the cycle $\gamma_j$, that confuse with $x_{jr}$. In this case, $|{\cal C}(w)|\neq 1$.

\coro
Let $G=(N,M_0,E,\ell)$ be an LPN system whose $T_u$-induced subnet is acyclic, $V=(\tilde{N},\tilde{M_0},E,\tilde{\ell})$ the VN of $G$, and $B = (X, E, f, x_0)$ the BRG of $V$.  LPN $G$ does not perform any detectability if for all cycles $\gamma_j$ in $B$, for all states
$x_{jr}\in \gamma_j$, $x_{jr}\notin X_m$.
\ecoro

\them\label{therm:PS-detect}
Let $G=(N,M_0,E,\ell)$ be an LPN system whose $T_u$-induced subnet is acyclic, $V=(\tilde{N},\tilde{M_0},E,\tilde{\ell})$ the VN of $G$, and $B = (X, E, f, x_0)$ the BRG of $V$.  LPN $G$ is periodically strongly detectable if and only if for all cycles $\gamma_j$ in $B$,
$\exists x_{jr}\in \gamma_j$, $x_{jr}\in X_m$.
\ethem

\prof
(If) Assume LPN $G$ is not periodically strongly detectable, that is for all $K\in\mathbb{N}, \exists \sigma\in L^\omega(G)$, there exist $\sigma' \preceq \sigma$, $\forall \sigma''\in T^*: \sigma'\sigma''\preceq \sigma, w'=\ell(\sigma'\sigma''), |\ell(\sigma'')|<K \wedge |{\cal C}(w')|\neq1$ in $G$. By Lemma~\ref{lem:seq-VN}, there must exist a sequence $\tilde{\sigma}\in L^\omega(V)$, $\tilde{\sigma'}\preceq \tilde{\sigma}$ and $\tilde{\sigma}''\in \tilde{T}^*$, with $\tilde{\ell}(\tilde{\sigma})=\ell(\sigma)$, $\tilde{\ell}(\tilde{\sigma'})=\ell(\sigma')$, $\tilde{\ell}(\tilde{\sigma''})=\ell(\sigma'')$.
Since $\tilde{\sigma}\in L^\omega(V)$ is of an infinite length and $B$ has a finite number of nodes, the path along $\tilde{\ell}(\tilde{\sigma})=w$ must contain a cycle $\gamma_j=x_{j1}e_{j1}x_{j2}\ldots x_{jk}\allowbreak e_{jk}x_{j1}\in \Gamma$, i.e., there exist $w_1\in E^*$ such that $w=w_1(e_{j1}\ldots e_{jk})^*$ and $|w_1|$ is finite.
Since the $T_u$-induced subnet of $G$ is acyclic, by proposition~\ref{prop:suf}, the $\tilde{T}_u$-induced subnet of $V$ is acyclic.
Thus let $\tilde{\ell}(\tilde{\sigma}_1)=w_1$, for all $\tilde{\sigma}_2\in \tilde{T}^*$ with $\tilde{\ell}(\tilde{\sigma}_2)=w_2 \preceq (e_{j1}\ldots e_{jk})^*$, $x_{jr}=f(x_0,w_1w_2)$.
Since $|{\cal C}(w')|\neq1$, $w'=\tilde{\ell}(\tilde{\sigma}'\tilde{\sigma}'')=w_1w_2$, thus $x_{jr}=f(x_0,w_1w_2)$ $\wedge$ $|{\cal C}(w_1w_2)|\neq1$.

\textbf{Case 1}:
If any $f(x_0,w_1w_2)=x_{jr}\notin X_m$. Therefore, $\gamma_j$ is the cycle that any state $x_{jr}\in \gamma_j$, $x_{jr}\notin X_m$.

\textbf{Case 2}:
If there is a state $f(x_0,w_1w_2)=x_{jr}\in X_m$.
let $w_2=(e_{j1}\ldots e_{jk})^*(e_{j1}\ldots e_{jr})$, since $f(x_0,w_1w_2)=x_{jr}\in X_m$ and $|{\cal C}(w_1w_2)|\neq1$, thus there must exist a state $x_{ir}$ in $B$ that $f(x_0,w_1w_2)=x_{ir}\neq x_{jr}$. By Algorithm~\ref{alo:xm}, $x_{ir} \notin \gamma_j$. Since $w_1w_2$ is of an infinite length and $B$ has a finite number of nodes, thus $x_{ir}$ must belong to another cycle $\gamma_i$ who have no intersection with $\gamma_j$. And the observation of $\gamma_i$ is $(e_{j1}\ldots e_{jk})^*$. Therefore, any state from $\gamma_j$ is confused with $\gamma_i$. According to the construction of VN and BRG, thus there must exist a cycle $\gamma_j'$ that any $x_{jr}'\in \gamma_j'$, $x_{jr}'\notin X_m$.


(Only if) Assume there exists a cycle $\gamma_j$ in $B$, that $\gamma_j=x_{j1}e_{j1}x_{j2}\ldots x_{jk}\allowbreak e_{jk}x_{j1}\in \Gamma: \forall x_{jr}\in \gamma_j$, $x_{jr}\notin X_m$.
Clearly, there exist $\tilde{\sigma}\in L^\omega(V)$ and $w_1\in E^*$ such that $\tilde{\ell}(\tilde{\sigma})=w=w_1(e_{j1}\ldots e_{jk})^*$ with $|w_1|$ is finite.
Since the $T_u$-induced subnet of $G$ is acyclic, by proposition~\ref{prop:suf}, the $\tilde{T}_u$-induced subnet of $V$ is acyclic.
Thus there exists $\tilde{\sigma_1}\preceq \tilde{\sigma}$ with $\tilde{\ell}(\tilde{\sigma}_1)=w_1$, for all $\tilde{\sigma}_2\in \tilde{T}^*$ with $\tilde{\ell}(\tilde{\sigma}_2)=w_2 \preceq (e_{j1}\ldots e_{jk})^*$, $f(x_0,w_1w_2)=x_{jr}$.
Since the sequence $\tilde{\sigma}\in L^\omega(V)$, $\tilde{\sigma_1}\preceq \tilde{\sigma}$ and $\tilde{\sigma}_2\in \tilde{T}^*$, by Lemma~\ref{lem:seq-VN}, there must exist a sequence $\sigma\in L^\omega(G)$ and exist $\sigma_1\preceq \sigma$ and ${\sigma}_2\in {T}^*$, with $\ell(\sigma_1)=\tilde{\ell}(\tilde{\sigma_1})=w_1$, $\ell(\sigma_2)=\tilde{\ell}(\tilde{\sigma_2})=w_2$.
By assumption, $x_{jr}\notin X_m$. Therefore, $f(x_0,w_1w_2)=x_{jr}\notin X_m$. By Proposition~\ref{prop:xm}, thus $|{\cal C}(w_1w_2)|\neq1$.
\eprof

In words, an LPN system is periodically strongly detectable if and only if any cycle in the BRG of its VN, that the cycle contains at least one marked state.

%

\begin{figure}
  \centering
  \includegraphics[width=0.35\textwidth]{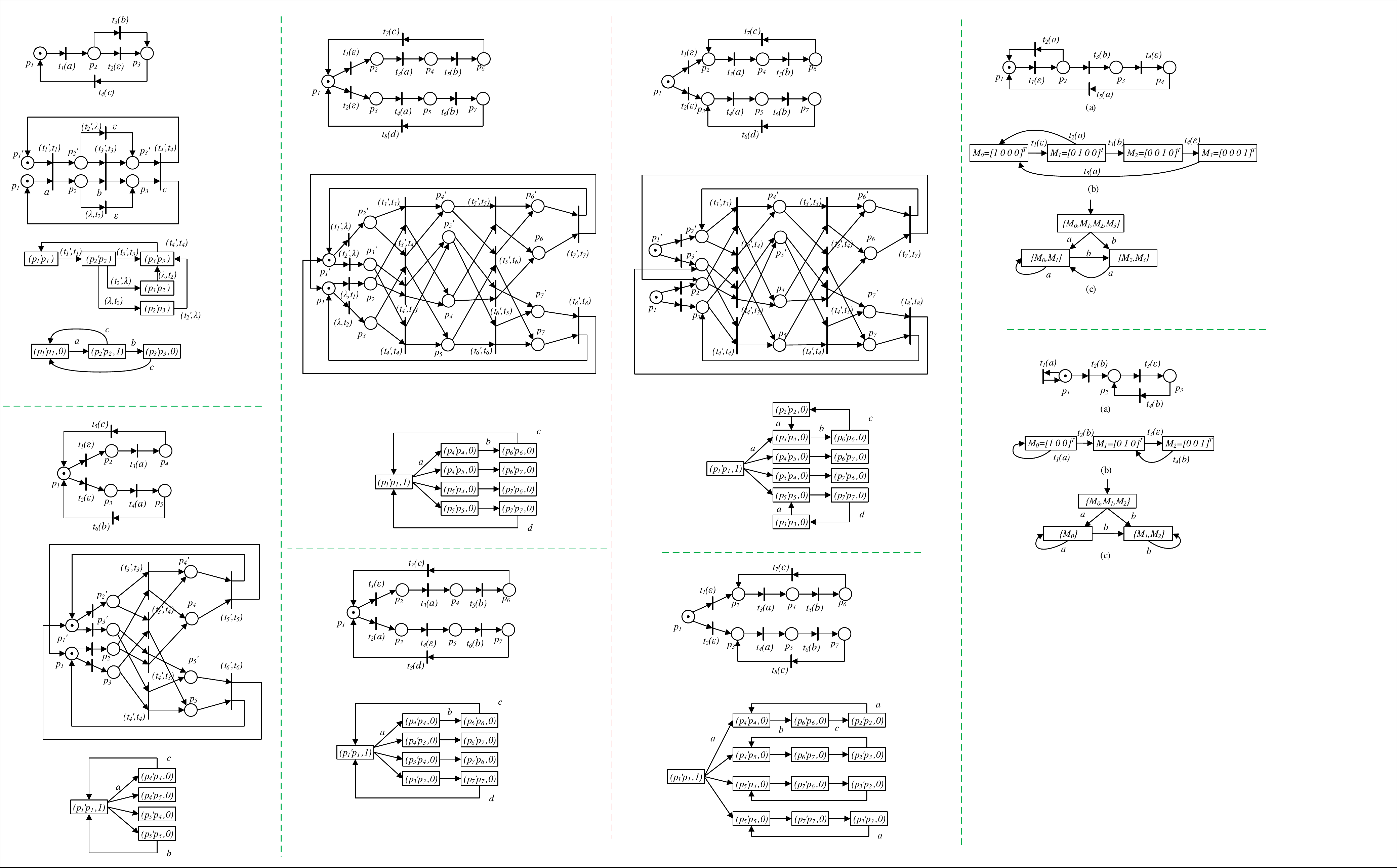}\\
  \caption{The LPN system in Example~\ref{eg:PSD}.}\label{fig:LPN2}
\end{figure}

\begin{figure}
  \centering
  \includegraphics[width=0.5\textwidth]{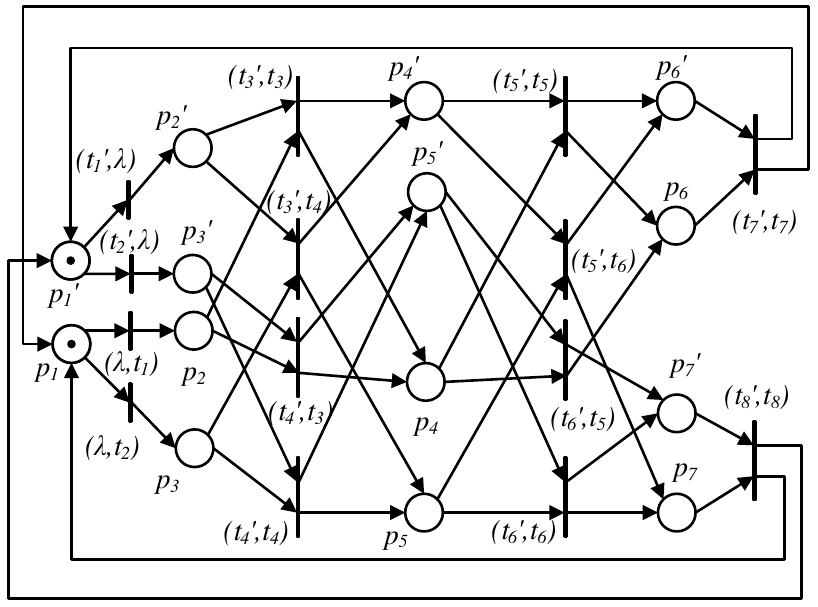}\\
  \caption{The VN system of the LPN in Fig.~\ref{fig:LPN2}.}\label{fig:VN2}
\end{figure}

\begin{figure}
  \centering
  \includegraphics[width=0.35\textwidth]{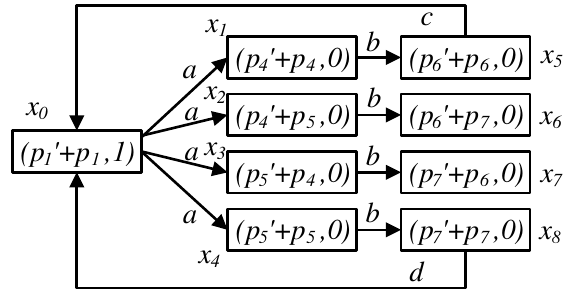}\\
  \caption{The BRG of the VN in Fig.~\ref{fig:VN2}.}\label{fig:BRG2}
\end{figure}

\exm\label{eg:PSD}
Now we use Theorem~\ref{therm:PS-detect} to check its periodically strong detectability.

\textbf{Case 1}: Consider again the LPN system in Fig.~\ref{fig:LPN1}. Its VN is shown in Fig.~\ref{fig:VN1}, and the BRG of the VN is shown in Fig.~\ref{fig:BRG1}. We know that there are tow simple cycles, $\gamma_{j1}=x_{1}bx_{5}ax_{1}$ and $\gamma_{j2}=x_{4}cx_{6}ax_{4}$.
As we get in Example~\ref{eg:SD}, $x_5(2)=1$ and $x_4(2)=1$, thus $x_5\notin X_m$, $x_4\notin X_m$. By Algorithm~\ref{alo:xm}, $X_m=\{x_{1}, x_{6}\}$.
Thus, in $\gamma_{j1}$ for observation $(ba)^*$, the current state $x_{1}$ is considered periodically distinguished.
And, in $\gamma_{j2}$ for observation $(ac)^*$, we can also periodically determine the current state $x_{6}$.
Therefore, by Theorem~\ref{therm:PS-detect}, the LPN system is periodically strongly detectable.

\textbf{Case 2}: Let us consider the LPN system in Fig.~\ref{fig:LPN2}. Its VN is shown in Fig.~\ref{fig:VN2}, and the BRG of the VN is shown in Fig.~\ref{fig:BRG2}. We can see there are two simple cycles, $\gamma_{j1}=x_{0}ax_{1}bx_{5}cx_{0}$ and $\gamma_{j2}=x_{0}ax_{4}bx_{8}cx_{0}$. For $x_0$, since $\alpha(\tilde{M_0})=1$, thus by Algorithm~\ref{alo:xm}, $x_0 \notin X_m$. For $x_1$, there exists an observation $w=bca$ that $f(x_5,w)=x_2\neq x_1$, Thus by Algorithm~\ref{alo:xm}, $x_1 \notin X_m$. The same for other states, by Algorithm~\ref{alo:xm}, we can get that $X_m=\phi$. Therefore, according to the Theorem~\ref{therm:PS-detect}, the LPN system is not periodically strongly detectable.
\hfill $\diamond$
\eexm

\subsection{Weak detectability and periodically weak detectability}\label{subsec:WD}

As we mentioned before, weak detectability and periodically weak detectability can not be verified by the BRG (even RG) of VN. It is because that when we construct the VN of LPN, $G'$ is a copy of $G$, thus we can always find a $\sigma'$ in $G'$ and a $\sigma$ in $G$ that for any $\sigma_1\preceq \sigma$ there exists a $\sigma_1'\preceq \sigma'$, $M_0'[\sigma_1' \rangle M'$, $M_0[\sigma_1 \rangle M$ with $\ell'(\sigma_1') = \ell(\sigma_1)$ and $M'=M$.
Thus, according to the construction of VN, if there exists a transition sequence $\sigma$ with no unobservable transition in $G$, then we can always find a path $\tilde{\sigma}$ in VN that for every marking $\tilde{M}$ in the path that $\tilde{M}=$
$\left[\begin{array}{cc}
  M'\\
  M
\end{array}\right]$
with $M' = M$.
Further, if the path $\tilde{\sigma}$ contain a cycle, by construction of BRG, that any state in the cycle of BRG can belong to $X_m$, even though the path may confuse with other path, i.e., there may exist $\tilde{\sigma'}$ that $\ell'(\tilde{\sigma'}) = \ell(\tilde{\sigma})$ and $\tilde{\sigma'}\neq \tilde{\sigma}$. And by the definition of the (periodically) weak detectability, we just need to check if there exists a cycle that is determined. Thus, for a subclass of LPNs, we can always find a cycle that all the states of the cycle are distinguishable in the BRG of its VN.

Therefore, the above situation would influence our verification on (periodically) weak detectability if we just analyze the BRG of the VN.

\begin{figure}
  \centering
  \subfigure[]{%
  \label{fig:LPN3}
  \includegraphics[width=0.2\textwidth]{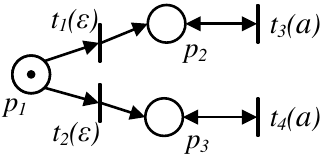}}
  \centering
  \subfigure[]{%
  \label{fig:VN3}
  \includegraphics[width=0.2\textwidth]{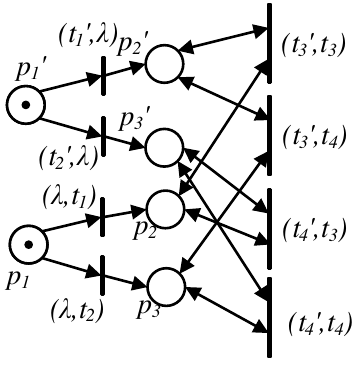}}
  \centering
  \subfigure[]{%
  \label{fig:BRG3}
  \includegraphics[width=0.3\textwidth]{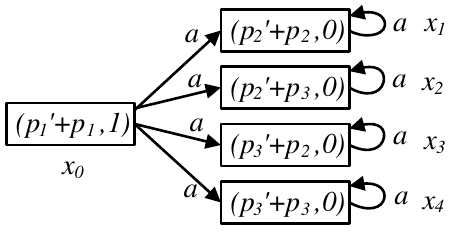}}

  \caption{The LPN system in Example~\ref{eg:WD} (a), the VN of the LPN (b), and the BRG of the VN (c).}
  \label{fig:WD}
\end{figure}

\exm\label{eg:WD}
Consider the LPN system in Fig.~\ref{fig:LPN3}. Its VN is shown in Fig.~\ref{fig:VN3}, and the BRG of the VN is shown in Fig.~\ref{fig:BRG3}. In the BRG, we can get that the cycle $\gamma_{j1}=x_{1}ax_{1}$ that the states in it can always be determined. Thus it seems that the LPN system is weakly detectable and periodically weak detectability. However, by the Definition~\ref{def:WD}, from the LPN system in Fig.~\ref{fig:LPN3}, we can easily know that it does not perform any detectability.
\hfill $\diamond$
\eexm

Although the weak detectability and periodically weak detectability can not be verified by the BRG of VN, it is easy to find that we can construct the observer of the BRG and the problem is able to be solved. However, for an LPN system, its four detectabilities can be verified by the observer of the LPN's BRG \cite{tong2019verification}. Thus, it is more complex if we through the observer of VN's BRG to check the detectability. Therefore the approach of VN is too complex for the weak detectability and periodically weak detectability, and we will not propose it.

\subsection{Computational complexity analysis}\label{subsec:complexity}

In this subsection we compare the computational complexity of the proposed approach with a previous approach in the literature \cite{tong2019verification}.

By Algorithm~\ref{alo:Verifier}, we can find that the number of places and tokens of a VN is twice that of the LPN. And in the worst case, the number of the transitions of a VN is $m^2$, where $m$ is the number of the transitions of the LPN. Thus, the complexity of constructing a VN is polynomial to the size of the original LPN system.
According to \cite{ziyue2017basis,tong2015verification}, in the worst case, the complexity of constructing a BRG is equal to that of the RG. Thus, in the worst case, the complexity of constructing the BRG is exponential to the VN's size.

In \cite{tong2019verification}, the detectability property of an LPN system can be decided by constructing the observer
of its BRG. In that approach, the complexity of constructing the BRG of the LPN is exponential to its size in the worst case. Moreover, in the worst case, the complexity of constructing the observer is also exponential to the number of states of the BRG.

Therefore, compared with the two approach, the proposed approach in this paper is more efficient in general.


\section{Conclusion}


In this paper, a novel approach to verifying detectability of bounded labeled Petri nets is developed. Our approach is based on the new tool called a verifier net, and on the exploration of its basis reachability graph for the detectability. For Petri nets whose unobservable subnet is acyclic, the strong detectability and periodically strong detectability property can be decided by just constructing the BRG of the verifier net. Since a complete enumeration of possible firing sequences is avoided and there is no need for the construction of observer, the proposed approach is of lower complexity than the previous approaches.
The future research is to study on an algorithm that can check the weak detectability and periodically weak detectability with low complexity.


\section*{Acknowledgment}

This work was supported by the National Natural Science Foundation of China under Grant No. 61803317, the Fundamental Research Funds for the Central Universities under Grant No. 2682018CX24, the Sichuan Provincial Science and Technology Innovation Project under Grant No. 2018027.

\ifCLASSOPTIONcaptionsoff
  \newpage
\fi



%


\bibliographystyle{IEEEtran}
\bibliography{verifier}

\end{document}